\documentclass[structabstract]{aa}  
\usepackage{graphicx}
\usepackage{dcolumn}
\usepackage{txfonts}
\usepackage{natbib}
\usepackage{natbib,twoopt}
\bibpunct{(}{)}{;}{a}{}{,} %% natbib format like A&A and ApJ
\newcommandtwoopt{\citeads}[3][][]{\href{http://adsabs.harvard.edu/abs/#3}%
{\citealp[#1][#2]{#3}}}
\newcommandtwoopt{\citepads}[3][][]{\href{http://adsabs.harvard.edu/abs/#3}%
{\citep[#1][#2]{#3}}}
\newcommandtwoopt{\citetads}[3][][]{\href{http://adsabs.harvard.edu/abs/#3}%
{\citet[#1][#2]{#3}}} 
\newcommandtwoopt{\citeyearads}[3][][]%
{\href{http://adsabs.harvard.edu/abs/#3}{\citeyear[#1][#2]{#3}}}

\begin{document}
\authorrunning{S. Mathur et al.}
\titlerunning{Magnetic activity with CoRoT and NARVAL}
 %  \title{Analysis of magnetic activity in solar-like stars observed by CoRoT and NARVAL}
   \title{Constraining magnetic-activity modulations in 3~solar-like stars observed by CoRoT and NARVAL}

   %\subtitle{I. Overviewing the $\kappa$-mechanism}

  \author{S. Mathur\inst{1,2}
          \and
          R.~A. Garc\'ia\inst{2}
          \and
          A. Morgenthaler\inst{3,4}%JB
          \and    
          D. Salabert\inst{5}%JB
          \and
	P. Petit\inst{3,4}%JB
          \and
          J. Ballot\inst{3,4}%JB
          \and 
         C. R\'egulo\inst{6,7}%JB
         \and
         C. Catala\inst{8}
          }
\institute{High Altitude Observatory, NCAR, P.O. Box 3000, Boulder, CO 80307, USA\\ \email{savita@ucar.edu}
\and
Laboratoire AIM, CEA/DSM-CNRS-Universit\'e Paris Diderot; IRFU/SAp, Centre de Saclay, 91191 Gif-sur-Yvette Cedex, France\\ \email{rgarcia@cea.fr}
\and
CNRS, Institut de Recherche en Astrophysique et Plan\'etologie, 14 avenue Edouard Belin, 31400 Toulouse, France\\ %JB
\email{amorgenthale@irap.omp.eu, ppetit@irap.omp.eu, jerome.ballot@irap.omp.eu} %JB
\and
Universit\'e de Toulouse, UPS-OMP, IRAP, 31400 Toulouse, France %JB
\and
Laboratoire Lagrange, UMR7293, Universit\'e de Nice Sophia-Antipolis, CNRS, Observatoire de la C\^ote d'Azur, BP 4229, 06304 Nice Cedex 4, France\\ \email{salabert@oca.eu}
\and
Instituto de Astrof\'isica de Canarias, E-38200 La Laguna, Tenerife, Spain\\
\email{crr@iac.es}
\and
Departamento de Astrof\'isica,  Universidad de La Laguna, E-38206 La Laguna, Tenerife, Spain
\and 
LESIA, CNRS, Universit\'e Pierre et Marie Curie, Universit\'e Denis Diderot, Observatoire de Paris, 92195 Meudon cedex, France
\email{claude.catala@obspm.fr}
}
  
  \date{Received August 18, 2011; accepted }

% \abstract{}{}{}{}{} 
% 5 {} token are mandatory
 
  \abstract
  {Stellar activity cycles are the manifestation of dynamo process running in the stellar interiors. They have been observed during years to decades thanks to the measurement of stellar magnetic proxies at the surface of the stars such as the chromospheric and X-ray emissions, and the measurement of the magnetic field with spectropolarimetry. However, all of these measurements rely on external features that cannot be visible during for example, a Maunder-type minimum. With the advent of long observations provided by space asteroseismic missions, it has been possible to pierce inside the stars and study their properties. Moreover, the acoustic-mode properties are also perturbed by the presence of these dynamos.}
 {We track the temporal variations of the amplitudes and frequencies of acoustic modes allowing us to search for signature of magnetic activity cycles, as has already been done in the Sun and in the CoRoT target HD~49933.} %JB
{We use asteroseimic tools and more classical spectroscopic measurements performed with the NARVAL spectropolarimeter to check if there are hints of any activity cycle in three solar-like stars observed continuously for more than 117 days by the CoRoT satellite: HD~49385, HD~181420, and HD~52265.
To consider that we have found a hint of magnetic activity in a star we require to have a change in the amplitude of the p modes that should be anti-correlated with a change in their frequency shifts, as well as a change in the spectroscopic observations in the same direction as the asteroseismic data. }
{Our analysis gives very small variation of the seismic parameters preventing us from detecting any magnetic modulation. However we are able to provide a lower limit of any magnetic-activity change in the three stars that should be longer than 120 days,  which is the length of the time series. Moreover we computed the upper limit for the line-of-sight magnetic field component being 1, 3, and 0.6 G for HD~49385, HD~181420, and HD~52265 respectively. More seismic and spectroscopic data would be required to have a firm detection in these stars.}
   {}

    %JB

   \maketitle
%
%________________________________________________________________

\section{Introduction}
The physical processes behind the dynamos producing magnetic activity cycles in stars are not yet perfectly explained \citep[e.g.,][and references there in]{2006ApJ...648L.157B,2010IAUS..264..120L}. Observing many stars showing magnetic cycles could help to better understand their dependence with the stellar properties and the place occupied by the Sun in this context. Moreover, the features at the surface of the stars, and in particular their magnetism, are extremely important to understand the characteristics of stellar neighborhoods and therefore, the properties and conditions in the exoplanets systems \citep[e.g.,][and references therein]{2010ApJ...714..384R}.

Stellar activity cycles have been measured for a long time \citep[e.g.,][]{1978ApJ...226..379W,1985ARA&A..23..379B,1995ApJ...438..269B,2007AJ....133..862H} mostly thanks to variations related to the presence of starspots crossing the visible stellar disk \citep[e.g.,][]{2009IAUS..259..363S}. Indeed, in many cases, these cycles were in a range from 2.5 to 25 years. Based on emission proxies, \citet{2002ESASP.485...35B} could estimate that about two-thirds of solar-type stars lie in the same range of magnetism as the Sun (minimum to maximum) with one-third of them being more magnetically active. Recently, spectropolarimetric observations of cool active stars unveiled the evolution of magnetic topologies of cool stars across their magnetic cycles, witnessed as polarity reversals of their large-scale surface field \citep{2009MNRAS.398.1383F,2009A&A...508L...9P}. More recently, other short activity cycles have also been detected using chromospheric activity indexes \citep{2010ApJ...723L.213M}. With all this information, it has been suggested that the length of the activity cycle increases proportionally to the stellar rotational periods along two distinct paths in main-sequence stars: the active and the inactive stars \citep{1999ApJ...524..295S,2007ApJ...657..486B}.

However, most of the activity-cycle studies are based on proxies of the surface magnetism at different wavelengths. This could be a problem as solar-like stars can suffer from periods of extended minima as it happened to the Sun during the Maunder minimum or between cycles 23 and 24. %JB2 debut
Nevertheless, during this unexpectedly long minimum, there were seismic evidences for the start of a new cycle in 2008 whereas the classical surface indicators were still at a low level \citep{2009A&A...504L...1S,2010ASPC..428...43F}.
{  Beside, \citet{1998SoPh..181..237B} showed that during the Maunder minimum, while no sunspots were visible on the surface, internal changes seemed to be going on. Therefore, having complementary diagnostics on the internal magnetic activity  of the stars could help to better understand the coupling between internal and external manifestations of magnetic phenomena.}

It is now well known that the frequencies of solar acoustic (p) mode change with the solar activity level \citep[see][]{1985Natur.318..449W,1989A&A...224..253P}. These changes in p modes are induced by the perturbations of the solar structure in the photosphere and just below \citep[e.g.,][]{1991ApJ...370..752G,2001MNRAS.324..910C}. 
Therefore, the variation with the magnetic cycle of the mean values of several global p-mode properties --with different geometrical weights-- will be sensitive to overall changes in the structure of the Sun and not to a particular spot crossing the visible solar disk.

The advent of long and continuous asteroseismic measurements, provided by the recent spacecrafts such as CoRoT  \citep[Convection Rotation and planetary Transits,][]{2006ESASP1306...33B,2008Sci...322..558M} or {\it Kepler} \citep[][for a description of the solar-like observations done with this instrument]{2010Sci...327..977B,2011Sci...332..213C}, allows us to study magnetic activity cycles by means of p-mode frequencies and amplitudes as has already been done in the Sun using Sun-as-a-star observations \citep[e.g.,][]{1992A&A...255..363A,2004ApJ...604..969J}. This technique has already been used successfully on HD~49933, a solar-like star observed by CoRoT \citep{2010Sci...329.1032G,2011A&A...530A.127S}. %JB2 fin
These same observations can also be used to study the surface magnetism with starspots proxies \citep{2010Sci...329.1032G,2011ApJ...732L...5C} or using spot-modeling techniques \citep[e.g.,][]{2009A&A...506...33M,2010A&A...518A..53M}.

%Since its launch, the CoRoT satellite has already observed several solar-like stars during periods longer than 117 days \citep[e.g.][]{2009A&A...507L..13B,2009A&A...506...41G} in the direction of both the galactic center and its anti-center. 

In this work, we study the temporal variations of the p-mode characteristics of three of the highest signal-to-noise ratio solar-like stars observed by CoRoT during more than 117 continuous days: HD~49385 \citep{2010A&A...515A..87D}, HD~181420 \citep{2009A&A...506...51B}, and HD~52265 \citep{2011A&A...530A..97B}. In Section 2 we describe the methodology followed in this paper to analyze both, the asteroseismic and spectroscopic observations. Then, in Section 3, we discuss the results obtained for each of the three stars studied. Finally, we give our conclusions in Section 4.

%Consequences/impact on dynamo: inputs of constraints for simulations
%Asteroseismology already used to measure magnetic activity (Ref to Science)
%Impact of short cycles: Ludwig et al.

\section{Methodology and data analysis}
\subsection{Asteroseismic parameters}

%Asteroseismic parameters . In the following sections, we describe how we compute the temporal variation of the frequency shifts of the acoustic modes, of the maximum amplitude of p modes and of the power spectra.

%\subsection{Frequency shifts}

Acoustic-mode parameters were obtained from the analysis of subseries of 30 days, shifted every 15 days,  therefore, only every other points are independent. We chose this length of subseries to have a good trade-off between the resolution in the acoustic modes and having enough subseries to study a hypothetical cycle. We had checked with solar observations (using the Global Oscillation at Low Frequency \citep{GabGre1995,GarSTC2005} instrument aboard the Solar Heliospheric Observatory) that the differences between different lengths of subseries are well within the uncertainties. In each of the subseries we used, individual p-mode frequencies were extracted with a maximum likelihood estimator \citep{1990ApJ...364..699A}. We fitted Lorentzian profiles with a local approach
on successive series of $l=0$, 1, and 2 modes \citep{2004A&A...413.1135S}. 

%{   The global modes are not sensitive to the presence of a spot on the stellar surface because the modes propagate in a large cavity. Moreover, we averaged several modes to compute the magnetic proxies ($A_{\rm max}$ or the frequency shifts), which means that we averaged different cavities inside the star. }

%A maximum likelihood estimator has been used because the observed power spectrum is distributed along the limit spectrum with a $\chi^2$ with 2 degrees-of-freedom statistics \citep{1990ApJ...364..699A}. 

The background was fitted using a Harvey-like profile describing the granulation \citep[e.g.,][]{1985ESASP.235..199H,2011ApJ...741..119M}. The amplitude ratio between the $l=0$, 1, and 2 modes was fixed to 1, 1.5, and 0.5 respectively \citep{2011A&A...528A..25S}, and only one linewidth was fitted per radial order.
For HD~52265, the rotational splitting and the inclination angle of the star were fixed to the values given in \citet{2011A&A...530A..97B}.
For the other two stars, HD~181420 and HD~49385, because of less reliable estimates in these parameters, no splitting and no inclination angle were fitted, i.e. only one Lorentzian profile was used to model each of the modes as it is commonly done in cases where the linewidth of the modes is too big to properly disentangle the splitting or when the signal-to-noise ratio is small \citep[e.g.,][]{2011ApJ...733...95M}. The mode identification \#1 from \citet{2009A&A...506...51B} and from \citet{2010A&A...515A..87D} were respectively used. Table~\ref{Tab1} summarizes the frequency range used to compute the averaged frequency shift of each star. It is important to note that we have also used cross-correlation techniques \citep{1989A&A...224..253P,2010Sci...329.1032G} to compute the frequency shifts in a global way. The results of both methods are quantitatively the same within their uncertainties. Hence, for the sake of clarity we have decided to show the frequency shifts obtained by the first method only. 

To extract the maximum amplitude per radial mode, $A_{\rm max}$, we  used the method described by \cite{2010A&A...511A..46M}, largely tested with the {\it Kepler} targets \citep{2010ApJ...723.1607H,2011A&A...525A.131H,2011MNRAS.415.3539V,2012A&A...537A..30M}. Briefly, we first subtract from the power density spectrum (PDS) the background model (one Harvey-law function as explained previously) fitted including a white noise component. 
We then smooth the PDS over 3$\times \langle \Delta \nu \rangle$, where $\langle \Delta \nu \rangle$ is the mean large separation obtained with the A2Z pipeline \citep{2010A&A...511A..46M}. Finally, we fit a Gaussian function around the p-mode bump giving us the maximum height of the modes. This power, $P_{\rm max}$, is converted into a bolometric amplitude, $A_{\rm max}$, by using the method based on \citet{2008ApJ...682.1370K} and adapted to CoRoT  by \citet{2009A&A...495..979M}, following the formula:

\begin{equation}
A_{\rm max}=\frac{4\sqrt{(P_{\rm max} \times \langle \Delta \nu \rangle)}}{R_{\rm osc}},
%§(4./Rosc)*sqrt(2*P_max*q*dnu/(q))/sqrt(2)
\end{equation}

\noindent {  where the response function $R_{\rm osc}$ for CoRoT observations given by \citet{2009A&A...495..979M} is:}

\begin{eqnarray}
R_{\rm osc}=7.134+(-96.8.10^{-5}) \times (T_{\rm eff}-5777) \nonumber \\
+ 13.10^{-8} \times (T_{\rm eff}-5777)^2
\end{eqnarray}

The frequency interval used in this analysis is defined in the last column of Table~\ref{Tab1}. We used different frequency ranges when we computed the frequency shifts and when we computed the maximum amplitude of the modes. Indeed, for the measurement of the frequency shift, we fit individual modes. We need a high signal-to-noise ratio to have lower uncertainties in the frequency shifts leading to a quite narrow frequency range. To measure the maximum amplitude per radial mode, we fit a Gaussian function over the p-mode bump and thus, generally, we need a much broader frequency range.

%It is slightly different from the range used for the computation of the frequency shifts because the fit is more stable when larger regions are used.

%{  Need to say the range use to fit the Gaussian. If different form Freq. Shifts, then add the ranges to Tab.1}

\begin{table}
\centering
\caption{Frequency range used to compute the averaged frequency shift and  
$A_{\rm max}$ of the 3 stars.}\label{Tab1}
\begin{tabular}{lccc}
\hline
\hline
Star &Frequency range for  & Frequency range for  \\ %& $\Delta \nu$  ($\mu$Hz) & $\nu_{\rm max}$ ($\mu$Hz) \\
 & frequency shifts ($\mu$Hz) & $A_{\rm max}$ ($\mu$Hz)\\
\hline
HD~49385 & 600--1300 & 550--1500\\% & 56.32 & 1013 \\ %JB2: - > --
HD~52265 & 1300--3200 & 1250--3700\\ %& 97.56 & \\ %JB2: - > --
HD~181420 & 900--2300 & 900--2300 \\ %& 75.35 & \\ %JB2: - > --

\hline
\end{tabular}
\end{table}

%{\bf We fitted the temporal variation of $A_{\rm max}$ and the frequency shifts to see if there was a quantitative relation between them. However the signal contained in these parameters is the sum of three components: the noise, the short-term variability and the long-term variability. We fitted all the data points of both parameters with a linear and second order polynomial function since we want to filter the short-term variability to search for global trends. The points used in this analysis are correlated because the shift in time between two consecutive points is half of the subseries length. To take that into account, we perform a weighted least-square fit \citep[e.g.][]{1992nrfa.book.....P}. This requires computing the variance-covariance matrix $\Sigma$ of the data.  For independent data points, $\Sigma$ is simply a diagonal matrix. In our case, every point is 50\% correlated with each of its direct neighbours. $\Sigma$ is then a tridiagonal matrix with the following components: $\Sigma_{i,j} = \sigma_i\sigma_j/2$ for $j=i+1$, $\Sigma_{i,i} = \sigma_i^2$, and zeros everywhere else ($\sigma_i$ denotes the error of the $i$-th point). The reduced $\chi^2$ is computed with the same matrix.  }

%\subsection{Starspot proxy}
For each star we also calculated a starspot proxy, as described in \citet{2010Sci...329.1032G} and \citet{2011ApJ...732L...5C}, i.e., we computed the standard deviation of the light curve of each subseries, which gives some information on the fluctuation of the brightness of the star. We assume that this modulation is due to starspots crossing the visible stellar disk. Indeed, the photometry of these CoRoT targets is dominated by the stellar signal and not by the photon noise. Taking as an example HD~52265, the measured level of the photon noise is $\sim$0.5 $\rm{ppm}^2/\mu$Hz \citep{2011A&A...530A..97B}, i.e. $\sim$\,88\,ppm in the flux, which is 3 to 4 times smaller than the dispersion we have measured with the magnetic proxy. {  However, the starspot proxy could be more perturbed compared to the global p modes. If a large spot crosses the visible stellar disk, on the one hand it will induce a more or less significant fluctuation in the light curve and thus in the starspot proxy. On the other hand, the global p modes will not be affected as they are sensitive to the global magnetic field of the star. As a consequence, there could be a mismatch between the evolution of the starspot proxy and the p modes.}

%\subsection{Surface rotation period}

To follow the time evolution of structures in the frequency domain all along the length of the observations, we analyzed the time series with the wavelet tool \citep{1998BAMS...79...61T, liu2007}. In our case, we took the Morlet wavelet, which is the product of a sinusoid and a Gaussian function and we calculate the wavelet power spectrum (WPS). By collapsing the WPS along the time axis, we obtain the global wavelet power spectrum (GWPS). With this technique, we are able to track the temporal evolution of the starspots and look for any increase in the activity level while we can verify the presence of any harmonic of the rotation at lower frequency. Indeed, the wavelet analysis reconstructs the signal putting most of the power in the fundamental harmonic reducing the leakage on the overtones. An example of the use of this technique to study the solar activity cycle and the rotation period of the Sun during the last 3 cycles (using a combination of real data and simulations) can be seen in \citet{2011JPhCS.271a2056V}. In that analysis it is shown how the main signature in the WPS is the rotation period at $\sim$\,26 days instead of the first overtone at  $\sim$\,13 days, which is the most important peak in the power spectrum of the Sun at low frequency for photometric and Doppler velocity observations (except during the minimum activity periods where no signal of the rotation can be measured above the general background level). We computed the 95\% confidence level for the GWPS to  quantify the detection of a peak. The 95\% confidence levels were obtained as described in Section 5 of \citet{1998BAMS...79...61T} knowing that the GWPS has a $\chi^2$ distribution.

\subsection{Spectroscopic analyses}

To complement the seismic studies, we observed our sample of three stars with the NARVAL spectropolarimeter placed at the Bernard Lyot, a 2m telescope at the Pic du Midi Observatory \citep{2003EAS.....9..105A}. The instrumental setup (in its polarimetric mode) and reduction pipeline are strictly identical to the one described by \citet{2008MNRAS.388...80P}. With the adopted instrumental configuration, it was possible to perform the simultaneous recording of a high-resolution spectrum in unpolarized and circularly polarized light. Except for a fraction of the data sets collected for HD~52265 and HD~181420, the available spectroscopic material is not contemporaneous to the CoRoT runs.

We used the unpolarized spectrum to estimate the stellar chromospheric emission in the cores of the \ion{Ca}{ii} H spectral line. We employed the same approach as the one used by \citet{2004ApJS..152..261W}, to calculate an activity proxy calibrated against Mount Wilson ÒS-indexÓ measurements.

For the Sun, the variations of the S index are $\sim$0.04 while the associated uncertainties in stellar measurements are typically of $10^{-3}$. Therefore the uncertainties are one order of magnitude smaller than the S-index variations for a star like the Sun. This sensitivity is enough to detect a stellar cycle similar to the one observed in the Sun. 

In a highly-sensitive search for Zeeman signatures generated by a large-scale photospheric field, we extracted from each polarized spectrum a mean photospheric line profile with enhanced signal-to-noise ratio, using the cross-correlation method as defined by \citet{1997MNRAS.291..658D}. The line-of-sight component of the magnetic field was then calculated with the centre-of-gravity technique \citep{1979A&A....74....1R}.

\begin{table}
\centering
\caption{Chromospheric emission in the core of the \ion{Ca}{ii} H spectral line.}\label{Tab2}
\begin{tabular}{lcc}
\hline
\hline
Star & S-index & ref.\\
\hline
HD~49385 & 0.139 & 1 \\
HD~52265 & 0.159 & 1 \\
HD~181420 & 0.245 & 1\\
%HD~49933 & 0.311  & 2 \\
Sun & [0.16--0.2]   & 3 \\ %JB2
\hline
\end{tabular}
\tablebib{(1) this work;  (2) [min--max] activity range, \citet{1995ApJ...438..269B}} %JB2
\end{table}
%(2)  Metcalfe, T.S. (private communication);

\section{Results}

\subsection{HD~49385}

HD~49385 is a G-type star, more evolved than the Sun, and thus placed in the HR diagram at the end of the main sequence or lying shortly after it. The effective temperature, $T_{\rm{eff}}$, is about  6095 $\pm$ 65 K, and it has a projected rotational velocity of $v \sin i = 2.9^{+1.0}_{-1.5}$ \,km\,s$^{-1}$. The seismic analysis of the CoRoT data provides a mean large spacing, $\langle \Delta \nu \rangle$\,=\,56.3\,$\mu$Hz, and the frequency of the maximum amplitude of the p-mode bump, $\nu_{\rm max}$, of 1013\,$\mu$Hz  \citep[see a detailed compilation of the parameters of the star in][]{2010A&A...515A..87D}. 

\begin{figure}[htbp]
\begin{center}
\includegraphics[width=7cm, angle=90, trim=0 1cm 0 1cm]{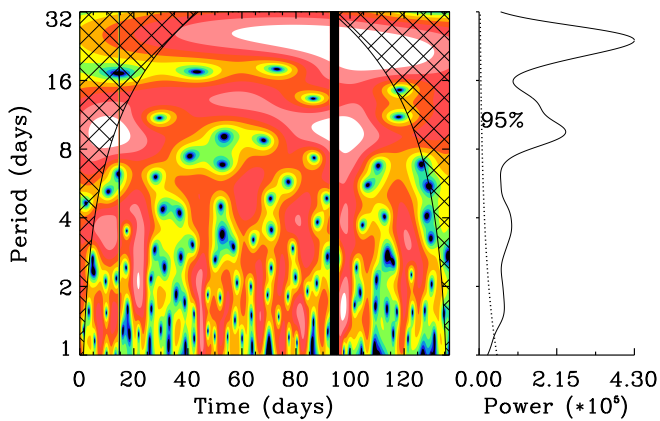}
\caption{Left panel: wavelet power spectrum (WPS) as a function of time of HD~49385. The origin of time is October 18, 2007. Right panel: global wavelet power spectrum (GWPS). The shaded region in the WPS, corresponds to the cone of influence, i.e. the region in which the periods cannot be analyzed due to the short length of the observations.}
\label{wave_hd49385}
\end{center}
\end{figure}

Unfortunately, for this star the analysis of the light curve or the rotational splittings of the acoustic modes neither provides with certainty the surface rotation period nor the internal rotation respectively. \cite{2010A&A...515A..87D} explain that there might be a hint of the rotation period, $P_{\rm rot}$,  at around 10 days. However, the WPS shown in Fig.~\ref{wave_hd49385},  does present some enhanced power around 29 days as well, with much higher power than the peak at  $\sim$\,10 days. Unfortunately,  the cone of influence, which shows the region where the WPS is reliable, is too close to this value and longer observations would be needed to confirm such periodicity. Note that this longer rotation period is compatible with the spectroscopic $v \sin i$ when we use the seismic parameters combined with the effective temperature given above and a radius of R=1.96\,$R_\odot$ obtained from the scaling relations based on solar values \citep{1995A&A...293...87K,2011ApJ...743..143H}. Indeed, with this longer rotation rate, a larger range of stellar-inclination angles (60\,$\pm$\,30\,$^\circ$) are allowed than if the rotation of 10 days is considered. A period of 29 days also agrees with the Skumanich law \citep{1972ApJ...171..565S}, which gives $\sim$\,27 days when assuming an age of 5\,Gyr as given from the model of \citet{2011A&A...535A..91D} as well as with the Asteroseismic Modeling Portal \citep{2009ApJ...699..373M,2012ApJ...749..152M}.

\begin{figure}[htbp]
\begin{center}
\includegraphics[width=7cm]{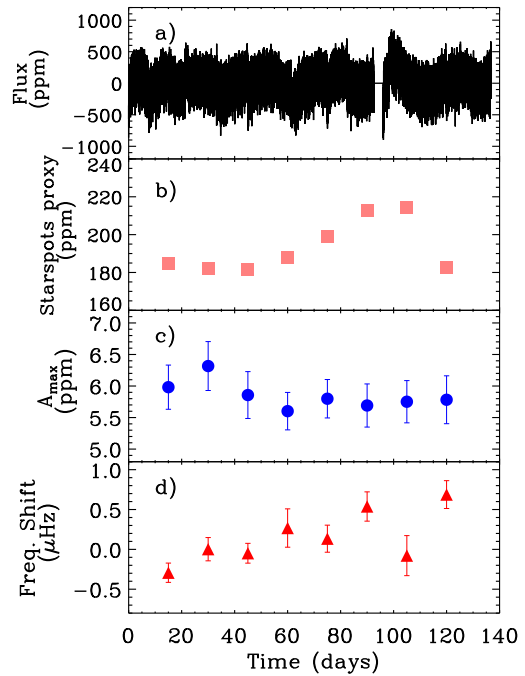}
\caption{({  a}): Flux of HD~49385 as a function of time (starting October 18, 2007) measured by CoRoT (only 1 point every 5 has been plotted). ({  b}): starspot proxy computed as described in Section 2. ({  c}): temporal variation of the maximum amplitude per radial mode with their associated error bar. ({  d}): temporal variation of the averaged frequency shift of the p modes. In panels b), c), and d), only every other point is independent. }
\label{hd49385}
\end{center}
\end{figure}

The temporal evolution of the maximum amplitude and of the averaged frequency shift of the acoustic modes are shown in Figs.~\ref{hd49385}~{  c} and {  d}.  

%{\bf The results of the first and second order fits are also represented by the dotted and the dashed line respectively.   The best-fit coefficients are listed in Table~\ref{Tab3}.   The small values of the slopes ($b$ coefficients), of the order of $10^{-2}$ and lower in absolute values with in general larger uncertainties}, could either indicate that the magnetic activity cycle is quite long or that the precision in the frequency shifts is too low to uncover a shorter or weaker cycle that would present smaller frequency shifts.  The reduced $\chi^2$ for $A_{\rm max}$ are smaller than 1, probably due to an overestimate of the uncertainties as the fits seem good on Fig.~\ref{hd49385}. Regarding the frequency shifts, the reduced $\chi^2$ are close to 10, probably because the short-term variations are not reproduced.

When comparing these trends visually, we see an anti-correlation between both parameters: $A_{\rm max}$, which slightly decreases; and the averaged frequency shift, which presents a slight increase during the observations.  According to what we know on the Sun and HD~49933, a situation like this corresponds to a small increase in the activity level of the star, which is corroborated by the surface activity measured by the starspot proxy (see Fig.~\ref{hd49385}~{  b}).  The correlation coefficient was computed using the SpearmanÕs rank correlation on independent points only (one over two, starting with the first one). We chose the rank correlation because the 2 variables are not expected to have a linear relationship. This analysis confirms that there is an anticorrelation between the two signals but the numerical value is small ($-$0.4, see Table~\ref{Tab3}).  The large value of the false-alarm probability associated to the correlation coefficient indicates that the anti-correlation is not firmly detected.

The spectropolarimetric observations of HD~49385 are constituted of 4 spectra collected between December 21, 2008  and April 13, 2009, i.e. one year after the CoRoT observations. Over this timespan, we observe  a low-level chromospheric flux, with an average value of the chromospheric index equal to 0.139 (see Table~\ref{Tab2}). The measured temporal fluctuations of the chromospheric emission are not statistically significant. The polarized spectra provide us with an upper limit of about 1 Gauss on the line-of-sight magnetic field component. This analysis is compatible with a general low-activity period. This result combined with the seismic observations  do not allow us to make a firm detection of any magnetic modulation in this star. However, HD~49385 will deserve to be revisited with CoRoT during the extension of the mission  to confirm or not the tendency unveiled here. It would be the first subgiant  with an on-going magnetic-activity cycle seismically observed.

\begin{table}
 \centering
 \caption{SpearmanÕs correlation coefficients of $A_{\rm max}$ and the averaged frequency shifts using only the independent points (one over two) starting by the first one and its associated false-alarm probability.
\label{Tab3}}
\begin{tabular}{l|cc}
\hline
\hline
Star        &  correlation coefficient   & False alarm probability\\
\hline

 HD~49385&  -0.4    & 48.9\%\\
 HD~52265&  0.4 & 48.9\% \\
  HD~181420&  0.3 & 59.9\%   \\

%----------------------------------------------- 
\hline
\end{tabular}
\end{table}

\subsection{HD~181420}

\begin{figure}[htbp]
\begin{center}
\includegraphics[width=7cm, angle=90, trim=0 1cm 0 1cm ]{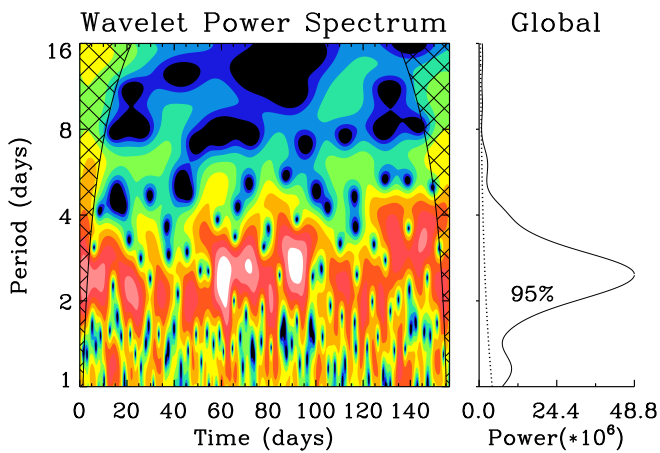}
\caption{WPS (left) and GWPS (right) of HD~181420 (same legend as in Fig.~\ref{wave_hd49385}). Observations started on May 11, 2007.}
\label{wave_hd181420}
\end{center}
\end{figure}

HD~181420 is an F2 star with $T_{\rm{eff}}\,=\,6580\,\pm\,105$\,K. The projected rotational velocity $v \sin i\,=\,18\,\pm\,1$\,km\,s$^{-1}$ \citep[see][for a detailed description of the spectroscopic properties of this star]{2009A&A...506..235B}. The surface rotation period, derived from the light curve observed by CoRoT, was estimated by \citet{2009A&A...506...51B} to be around 2.6 days with an important surface differential rotation. This yields -- in combination with the $v \sin i$ given previously -- an inclination angle for the star of  $35 \pm 21^\circ$. With the wavelet analysis, we also find this main periodicity (Fig.~\ref{wave_hd181420}). The WPS shows a broad peak confirming the presence of an important differential rotation at the surface of the star. Moreover, there is no signature of any other harmonic at longer periods.

\begin{figure}[!htbp]
\begin{center}
\includegraphics[width=7cm]{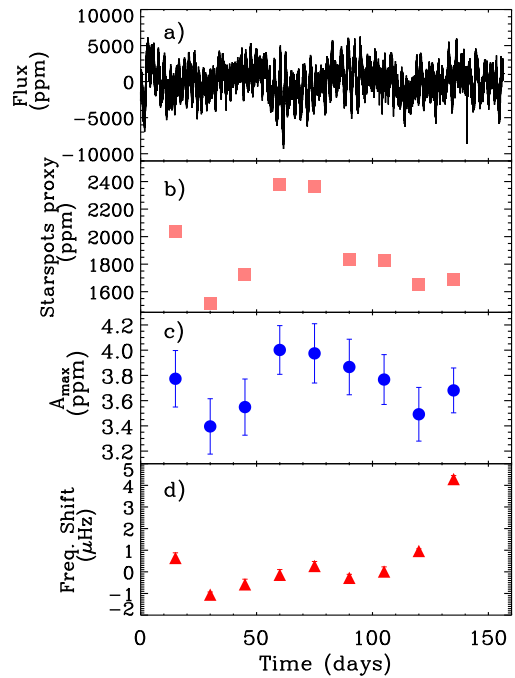}
\caption{Analysis of HD~181420. Same legend as in Fig.~2. Observations started on May 11, 2007. In panels b), c), and d), only every other point is independent.  } %JB
\label{hd181420}
\end{center}
\end{figure}

The temporal evolution of the averaged frequency shift and $A_{\rm max}$ are shown in Fig.~\ref{hd181420}~{  c} and {  d}.    $A_{\rm max}$ shows a small modulation while the averaged frequency shift exhibits a flat behavior (excepting the last point). Without surprise, the correlation of the independent points of the two signal is small with the same sign, thus a small correlation of 0.3 and a false-alarm probability close to 60\% (see Table~\ref{Tab3}).

When we analyze the light curve, in both the WPS and the starspots proxy, we see an increase in the signal around the sixtieth and the eightieth day of the measurements. This region of higher surface activity corresponds to the small maximum we observe in $A_{\rm max}$, which is not what we expect if this is due to an activity effect. We can conclude from the seismic analysis that we do not see any change related to a magnetic activity modulation in this star, while in the light curve we do see the presence of starspots (see Fig.~\ref{hd181420}~{  a}).

Spectropolarimetric data for HD~181420 were collected between June 2, 2007 and July 15, 2008 and overlap slightly with the asteroseismic data. Over this period, the average chromospheric index is equal to 0.245, with values ranging from 0.227 and 0.265 (see Table~\ref{Tab2}). The observed temporal fluctuations do not reveal a long-term trend. %JB (see Fig.~\ref{spec181420}). 
In spite of a chromospheric flux higher than solar, the signal-to-noise ratio of the polarized spectra is not sufficient for the detection of Zeeman signatures because of a significant rotational broadening of the line profile. We infer an upper limit of about 3 Gauss on the line-of-sight magnetic field component.

\subsection{HD~52265}

HD~52265 is a G0V, metal-rich, main-sequence star hosting a planet \citep{2000ApJ...545..504B,2001A&A...375..205N}. Its effective temperature is $T_{\rm eff}$\,=\,6100\,$\pm$\,60\,K and the projected rotational velocity is $v \sin i\,=\,3.6^{+0.3}_{-1.0}$\,km\,s$^{-1}$ \citep[see][for a complete review of the stellar parameters and the seismic analysis]{2011A&A...530A..97B}. Using the measurements provided by CoRoT -- during 117 continuous days starting on November 13, 2008 -- the rotation rate of the star was determined: $P_{\rm rot}$\,=\,12.3\,$\pm$\,0.15 days, with a stellar inclination angle of 30\,$\pm\,10^\circ$. 

\begin{figure}[!htbp]
\begin{center}
\includegraphics[width=7cm]{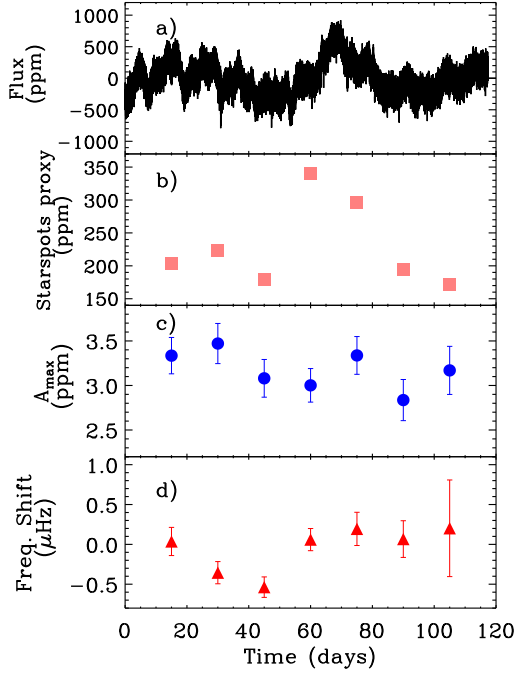}
\caption{Analysis of HD~52265. Same legend as in Fig.~2. Observations started in November 13, 2008.  In panels b), c), and d), only every other point is independent. }
\label{hd52265}
\end{center}
\end{figure}

  The visual inspection of the averaged frequency shift and $A_{\rm max}$ -- plotted in Fig.~\ref{hd52265} c and d --  presents very small variations. This is confirmed by the cross correlation coefficient which is of the order of +0.4, indicating that both signals are not anticorrelated (see Table~\ref{Tab3}). Once again the non-negligible false-alarm probability does not support the correlation between the two quantities.  
%as indeed the fits do not completely reproduce the data specially because of the point located at 45 days, which is lower than the other points.

%{\bf Visually, we observe a small anti-correlation between the amplitude variation and the frequency shifts.}

%the result for HD~52265 is very similar to HD49385 with a correlation coefficient of -0.64  with 88.1\% confidence level, suggesting an anti-correlation between $A_{\rm max}$ and the frequency shift. }

The starspots proxy (Fig.~\ref{hd52265}~{  b}) presents an increase around the middle of the data set. We notice that in the light curve displayed in Fig.~\ref{hd52265}~{  a}, there is indeed a sudden increase of the flux around that period. Although we cannot rule out a stellar origin, this kind of modulation is often due to instrumental instabilities in the CoRoT satellite and should then be treated with some precaution.  

\begin{figure}[!htbp]
\begin{center}
\includegraphics[width=\linewidth]{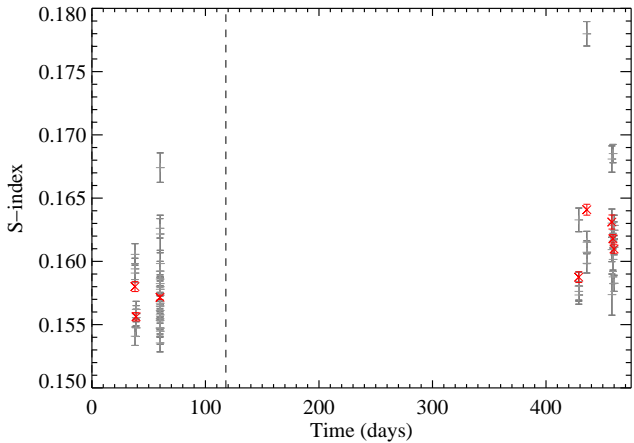} %JB
\caption{Chromospheric emission of the \ion{Ca}{ii} H line measured at different epochs with NARVAL for HD~52265. Grey points are the individual measurements while the red ones correspond to the daily averages. The starting date of the time series coincides with the beginning of CoRoT observations, i.e. November 13, 2008. The end of CoRoT observations is delineated with a vertical dashed line.} %JB
\label{spec52265}
\end{center}
\end{figure}

Our spectropolarimetric observations of HD~52265 were obtained between December 20, 2008 and January 11, 2009, i.e. the first part was collected during the CoRoT observations. From these time series, we derive an average chromospheric index of 0.159 (see Table~\ref{Tab2}), with a slight increase over the observing run, which is visible on Fig.~\ref{spec52265}. %JB2
Here again, the noise level in the polarized spectra remains too high to reach the detection threshold of Zeeman signatures, which can at least allow us to place a tight upper limit of 0.6\,Gauss on the longitudinal component of its large-scale photospheric field.

\section{Discussion and Conclusion} 

 In the present paper we have performed an extended analysis of seismic and spectroscopic data using CoRoT and NARVAL to look for signatures of magnetic activity cycles in three stars: HD~49385, HD~181420, and HD~52265. The seismic analysis was done by studying the temporal variation of the maximum amplitude of the modes, $A_{\rm max}$ and of the averaged frequency shifts, while the spectroscopic analysis consisted of measurements in Ca H chromospheric emission lines and we also researched for any Zeeman polarised signatures.

The information that we can extract from this analysis is limited by the data we have in hand {\bf and the length of the obser}vations. Indeed, we only have seven or eight data points and in each case, two consecutive points are correlated because of an overlap between the subseries analysed. 

Given the limitation of our analysis related to the length of the observations, we require that three criteria are fulfilled to be able to firmly detect magnetic modulations with our analysis. The amplitude of the modes and their frequency shifts must show a temporal variation. These two parameters have to be anti-correlated. Finally, we should observe a temporal variation in the spectroscopic data correlated with the seismic quantities.

For all the stars the seismic indicators show very small variations, that are, in general compatible with no variation at all at  a 2-$\sigma$ level. Only HD~49385 presents an anti-correlation between the two parameters, suggesting a rising phase of a possible magnetic activity cycle but with a small correlation coefficient ($\sim$- 40~$\%$) and a false-alarm probability of $\sim$\,50\%. No confirmation could be done with the spectropolarimetric measurements. Nevertheless we infer an upper limit of the line-of-sight magnetic field component of 1~G. Therefore, these results prevent us from concluding that we detected any magnetic modulation in this star. 

For the other two stars, HD~181420 and HD~52265, the correlation between the two seismic indicators are of positive sign and small, ruling out the existence of a magnetic modulation at this time scale. However, HD~52265 shows a small increase of the S-index (using two sets of observations separated    about one year). This could indicate a hint of a magnetic activity change towards a maximum of activity. However, as we do not have any variation in the seismic indicators we cannot claim for any firm detection of any magnetic activity change. The upper limit of the line-of-sight magnetic field component obtained was 3 and 0.6 G for HD~181420, and HD~52265 respectively.

In conclusion, the analysis presented here established that if there is a magnetic activity cycle in these stars it would be  longer than the CoRoT observation period of $\sim$ 120 days, which is a limiting factor. The false-alarm probabilities of the correlation coefficients are too large to confirm any trend. Moreover, through the spectroscopic analysis we can say that for HD~52265, the lower limit could be of about a year, but this should be taken carefully because the magnitude of the variation measured is very small.

 Longer datasets will be necessary to further investigate the presence of magnetic cycles in stars other than the Sun. This could be soon possible thanks to the data collected by the {\it Kepler} satellite or by revisiting these stars with CoRoT in the incoming years during the extension of the mission.

\bibliographystyle{aa} 
\bibliography{./BIBLIO_sav}

\begin{acknowledgements}
 The CoRoT space mission has been developed and is operated by CNES, with contributions from Austria, Belgium, Brazil, ESA (RSSD and Science Program), Germany and Spain. NARVAL is a collaborative project funded by France (R\'egion Midi-Pyr\'en\'ees, CNRS, MENESR, Conseil G\'en\'eral des Hautes Pyr\'en\'ees) and the European Union (FEDER funds). DS acknowledges the financial support from CNES. JB, RAG, SM, AM, and PP  acknowledge the support given by the French ``Programme National de Physique Stellaire''. RAG also acknowledges the CNES for the support of the CoRoT activities at the SAp, CEA/Saclay. This research has been partially supported by
the Spanish Ministry of Science and Innovation (MICINN) under the
grant AYA2010-20982-C02-02.NCAR is supported by the National Science Foundation. Wavelet software was provided by C. Torrence and G. Compo, and is available at URL: http://paos.colorado.edu/research/wavelets/
 \end{acknowledgements}

\end{document}